\documentclass[preprint,review,12pt,3p]{elsarticle}

\usepackage{etoolbox}

\usepackage{setspace}

\usepackage{amssymb}
\usepackage{url}
\usepackage{caption}
\usepackage{verbatimbox}

\usepackage{amssymb}
\usepackage{array,tabularx}
\usepackage{amsmath}
\usepackage{pgfplots}
\usepackage{graphicx}
\usepackage{color,soul}
\usepackage{amsthm}
\newtheorem{theorem}{Theorem}[section]

\newtheorem{lemma}[theorem]{Lemma}
\newtheorem{assumption}{Assumption}
\newenvironment{conditions*}
  {\par\vspace{\abovedisplayskip}\noindent
   \tabularx{\columnwidth}{>{$}l<{$} @{${}={}$} >{\raggedright\arraybackslash}X}}
  {\endtabularx\par\vspace{\belowdisplayskip}}

\usepackage{subcaption}
\usepackage{amsmath}
\usepackage{algorithm}

\usepackage[noend]{algpseudocode}

\usepackage{algpseudocode}
\algdef{SE}[DOWHILE]{Do}{doWhile}{\algorithmicdo}[1]{\algorithmicwhile\ #1}%
\DeclareMathOperator*{\argmin}{argmin} 
\DeclareMathOperator*{\argmax}{argmax} 

\journal{Journal of Parallel and Distributed Computing}

\pgfplotsset{compat=1.14}

\newenvironment{frontmatter2}{}{\pprintMaketitle}

\begin{document}

\begin{frontmatter2}

\title{An Efficient Fault Tolerant Workflow Scheduling Approach using Replication Heuristics and Checkpointing in the Cloud}


\author[1]{Amrith Rajagopal Setlur\corref{ref1}}
\ead{asetlur@cs.cmu.edu}
\cortext[ref1]{equal contribution (order based on coin flip)}

\author[1]{S. Jaya Nirmala\corref{ref1}}
\ead{sjaya@nitt.edu}

\author[1]{Har Simrat Singh}
\ead{hrsmrtsingh96@gmail.com}

\author[1]{Sudhanshu Khoriya}
\ead{sudhanshukhoriya@gmail.com}


\address{\textnormal{$^1$National Institute of Technology}}
\address{\textnormal{Tiruchirappalli - 620 015}}
\address{\textnormal{India}}

\begin{abstract}
Scientific workflows have been predominantly used for complex and large scale data analysis and scientific computation/automation and  the need for robust workflow scheduling techniques has grown considerably. But, most of the existing workflow scheduling algorithms do not provide the required reliability and robustness. In this paper, a new fault tolerant workflow scheduling algorithm that learns replication heuristics in an unsupervised manner has been proposed. Furthermore, the use of light weight synchronized checkpointing enables efficient resubmission of failed tasks and ensures workflow completion even in precarious environments. The proposed technique improves upon metrics like Resource Wastage and Resource Usage in comparison to the Replicate-All algorithm, while maintaining an acceptable increase in Makespan as compared to the vanilla Heterogeneous Earliest Finish Time (HEFT).

\end{abstract}


\end{frontmatter2}



\newpage

\section{Introduction}
\label{intro}

\par Scientific workflows are described as a "useful paradigm to describe, manage, and share complex scientific analysis" \cite{taverna}. A workflow is a formal way to express a calculation. The workflow involves multiple tasks of different sizes and characteristics, with control and data dependencies between them. They also capture the various parameters of the task such as their input, output etc.\cite{videoworkflow}. Workflows have emerged as comprehensive tools for managing complex computations and managing storage requirements. They are used in a lot of applications like neuroscience, high-energy physics and genetics. 

 Many authors have studied the advantages of using the cloud environment for executing scientific workflows \cite{hoffa2008use,deelman2010grids,juve2012resource} and have claimed that the cloud environment enables workflow execution with low cost and that the virtualization overhead due to the cloud would be very minimal (in the range of 1\% to 10\%). 
 \begin{sloppypar}
\par Efficient scheduling of scientific workflows helps in reducing the makespan or execution time, in meeting deadlines and in minimizing the cost. As the problem of scheduling tasks simultaneously on multiple processors with a start and an end time is NP-complete, researchers have relied on heuristic and meta-heuristic optimization techniques to schedule them.

 Failures in scientific workflows increase the makes\-pan and waste a lot of workforce and time. The different types of failures that can occur during an execution are task failures, VM failures and workflow-level failures. Fault tolerance for scientific workflows can be provided either at the task-level or workflow-level \cite{hwang2003grid}. Task-level scheduling techniques involve retry, checkpointing and the use of alternate resources for the same task. On the other hand, workflow level scheduling involves the usage of alternate tasks, redundancy, user-defined exception handling  and rescue workflows.  In this paper, a new fault tolerant workflow scheduling approach called Checkpointing and Replication based on Clustering Heuristics (CRCH) is proposed. It uses replication, resubmission, checkpointing and provides fault-tolerance in an efficient manner.
 \end{sloppypar}
  
 \par In the scheduling step, the workflow tasks are replicated and then scheduled. The multiple copies prevent the task from failing and increase the probability of its successful completion. If one copy fails, one of its replicas is scheduled and executed \cite{plankensteiner2009new, zhang2009combined}. Task resubmission \cite{grandstrand:2004,plankensteiner2009new} is also widely used for fault tolerance in the workflow scheduling. It takes place during the execution phase. In task resubmission, the failed task is resubmitted either to the same or different resource. The resource usage and wastage in task resubmission is less as compared to replication but the execution time may be more. Replication generates identical copies of a task. Hence they have same dependencies, and thus sufficient parallel systems can afford to execute them in parallel saving execution cost. Checkpointing \cite{grandstrand:2004} \cite{zhang2009combined} is one of the time efficient fault tolerant methods. In synchronous checkpointing, the states of the tasks or processes are saved promptly at regular intervals. Whenever a  Virtual Machine (VM) fails, the process starts from the previously saved state. Thus, this method is gainful over methods which reschedule a duplicate of the task.
\par One of the key contributions of this paper is an unsupervised way of learning replication counts for tasks. In comparison to other replication heuristics \cite{plankensteiner2009new},  this approach is much quicker and robust, as it doesn't involve exploring every possible solution (HEFT schedules with varying sets of replicas) in a combinatorial optimization problem. Along with this, a checkpointing mechanism that stimulates dynamic resubmission of tasks on the most optimum resource has been proposed. In an elaborate analysis of well established metrics like Resource Usage, Resource Wastage and Total Execution Time it has been shown that the algorithm proposed performs better than the existing state-of-the-art workflow scheduling techniques even in highly faulty environments. 
    \par The outline of the paper is as follows: \textit{Section \ref{relatedwork}} discusses some of the novel and significant progress made in the field of fault tolerant workflow scheduling. \textit{Section \ref{proposed_methodology}} and \textit{Section \ref{performance_analysis}} discuss in depth the algorithms we propose, followed by a performance benchmarking against state-of-the-art methods. The concluding remarks along with future research is presented in \textit{Section \ref{conclusion}}.
    
\section{Related Work}
\label{relatedwork}

\begin{sloppypar}
Yu and Buyya \cite{yu2005taxonomy} give a brief overview of the various fault tolerant workflow scheduling techniques. Fault tolerance to workflow applications are provided either at the task or workflow level. \textit{Replication} of tasks or data, \textit{Resubmission}, \textit{Checkpointing} and \textit{Alternate Resource} are widely used techniques at the task-level, whereas \textit{Alternate Task}, \textit{User-defined Exception Handling} and \textit{Rescue Workflow} are widely used at the workflow level. 
Poola et al. \cite{grandstrand:2004} give a comprehensive survey of the fault tolerant techniques employed in various Workflow Management Systems (WFMS). They also present a detailed taxonomy of the different techniques employed for fault tolerance in distributed environments. Also, the paper discusses a variety of metrics used for quantifying fault tolerance. \cite{plankensteiner2007fault} discusses the fault-tolerant techniques employed in various grid WFMS. The survey reveals that resubmission techniques are most widely used for providing fault tolerance in workflows followed by replication and checkpointing. 
\end{sloppypar}
\begin{sloppypar}
Poola et al. \cite{poola2014robust} use the concept of \textit{slack time} to generate robust schedules for scientific workflows to enable them to withstand failures in the cloud environment. They use a common set of parameters to model the stochasticity of all VMs. They also assume that there exists no resource contention, which can be a strong assumption in highly faulty environments. In this proposed methodology, resource failure parameters are sampled from various distributions, thus making the system more robust. Plankensteiner et al. \cite{plankensteiner2009new} estimate the replication count of a task from its Resubmission Impact (RI) heuristic. Their approach creates multiple workflows, each with a particular task duplicated by a constant value. The replication count is estimated from a normalized score assigned to each task, based on how much they impact the execution time (had they been replicated). On the other hand, the approach in this paper infers replication count using an unsupervised machine learning algorithm that gives more accurate estimates and saves the time involved in re-computing HEFT schedules.

Zhang et al. \cite{zhang2009combined} integrate the vanilla HEFT \cite{topcuoglu2002performance}/ Duplication Scheduling Heuristic (DSH)\cite{kruatrachue1988grain} schedules with the over-provisioning algorithm proposed by Kandaswamy et al.\cite{kandaswamy2008fault}. To meet the constraints on the overall workflow DAG (Directed Acyclic Graph) success probability, which diminishes exponentially with the addition of tasks to the workflow, the entire DAG is over-provisioned on a distinct set of resources. This leads to increased Resource Usage. The over-provisioning algorithm proposed by \cite{kandaswamy2008fault} finds the solution for a combinatorial optimization problem, which meets both performance and reliability constraints for a task. Although the assumption of independent binomial distributions for resource failures seems reasonable, the computation of expected execution time of a task on a resource cannot be agnostic to the state of the current workflow execution.

\end{sloppypar}
\begin{sloppypar}
Chen and Deelman \cite{chen2012fault} introduce horizontal/vertical task clustering based on the workflow structure. Having defined Gamma/Weibull distributions, for task runtime, overhead time and job (collection of tasks) runtime, they use Maximum Likelihood Estimates (MLE) for the distribution parameters, along with corresponding conjugate priors for dynamic estimation. For the parameter estimation to converge, large datasets of task/job runtimes are required. But the clustering technique proposed in this paper relies on grouping dense task embeddings. These embeddings are based on task/task-neighborhood structural characteristics (like edges, DAG order, etc.)
\end{sloppypar}

Zhang et al. \cite{zhang2009combined} use the technique to find the smallest subset of resources to replicate the tasks such that they satisfy their performance and reliability constraints. If the smallest subset of resources could not be found, the success probability for all the resource combinations are calculated, and the tasks are replicated on the resource set with highest success probability. The method proposed by \cite{plankensteiner2009new} does not use checkpointing, but resubmits a task when all of its replicas have failed. Resubmission of the whole task significantly increases the execution time of the task, which in turn increases the workflow's makespan. But the fault tolerant approach proposed in this paper employs replication heuristics and light-weight checkpoint/ restart techniques at the task level. The replication heuristic employed calculates the number of replications needed for each of the tasks in the workflow, and thereby reduce the resource waste and execution cost. Light-weight checkpointing enables the system to have minimal stable storage, and the transfer of intermediate data more manageable, and hence reduces execution time. 

\par Other popular checkpoint/restart algorithms include the Scalable Checkpoint/Restart (SCR) \cite{scr_checkpoint} and the FTI \cite{fti_checkpoint} algorithms. SCR is an asynchronous multi-level checkpointing algorithm that employs a combination of frequent inexpensive checkpoints and resilient time consuming backups to the Parallel File System (PFS). The checkpoint overhead of the vanilla SCR algorithm has been compared with CRCH in the latter half of the paper.

\section{Proposed Methodology}
\label{proposed_methodology}
\begin{figure*}
\centering
\includegraphics[width=0.90\textwidth, height=20cm, keepaspectratio]{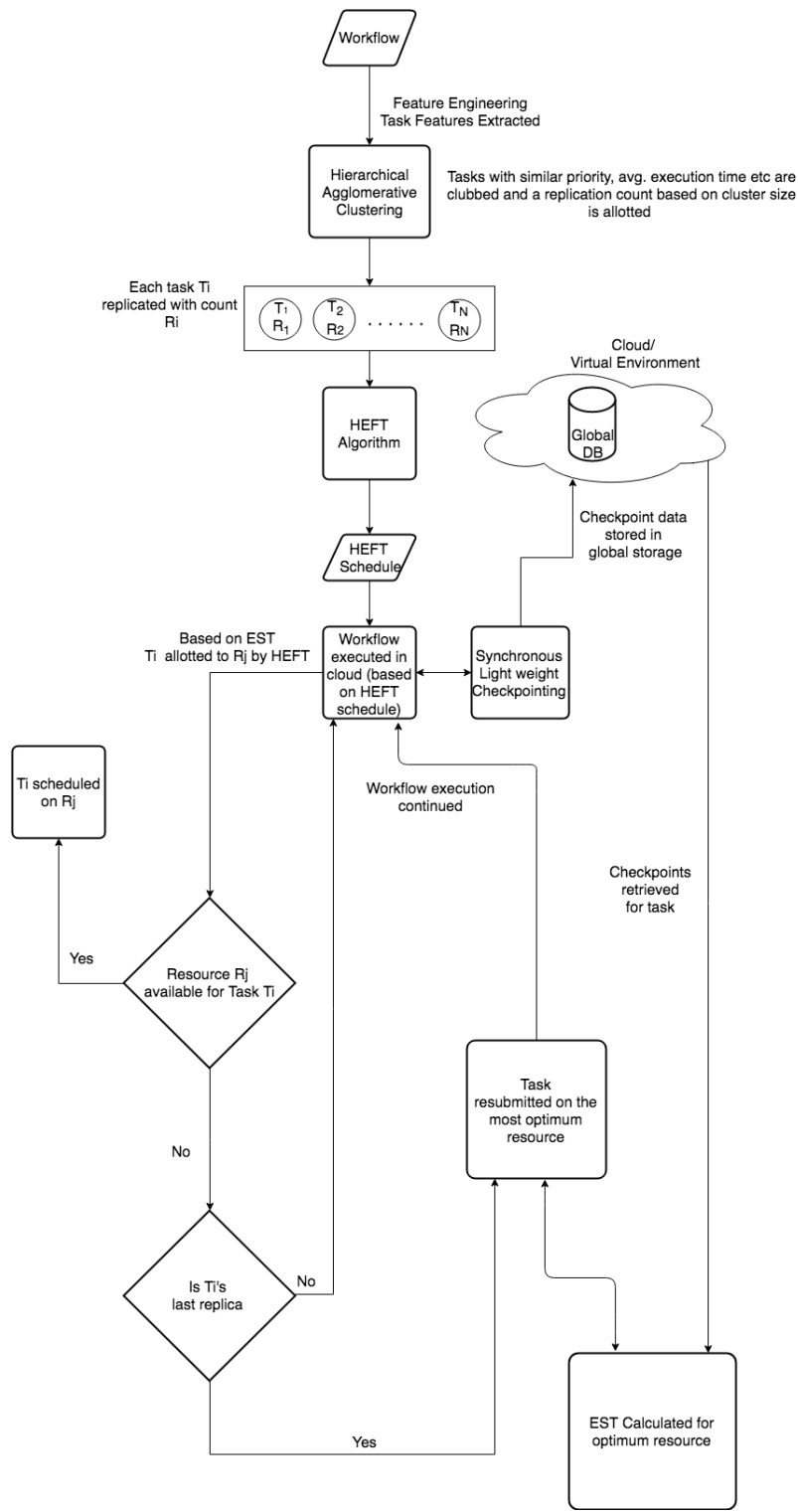}
\caption {CRCH Algorithmic Design}
\label{flow_chart}
\end{figure*}

\begin{table}[htp]
\caption{Notation Index}
\label{tab:1}       
\begin{tabular}{ll}
\hline{\smallskip}
Notation & Explanation  \\
\noalign{\smallskip}\hline\noalign{\smallskip}
$EST_t$ & Estimated Start Time of task $t$ \\
$EFT_t$ & Estimated Finish Time of task $t$ as decided by Algorithm-\ref{heftOP}.\\
     $AST_t$ & Actual Start Time of task $t$ \\
     $AFT_t$ & Actual Finish Time of task $t$ \\
     $TET$ & Total Execution Time of the Workflow \\
     $w_t$ & Average execution time of task $t$\\
     $e(t,t',d)$ & Average time to transfer $d$ units of data from $t$ to $t'$\\
   $timeOnVm(t,r)$ &Time taken for task $t$ to execute on resource $r$\\
     $dataTransfer(r,r')$ & Data transfer rate between $r$ and $r'$\\
  $taskList$ & List of tasks in workflow\\
     $vmList$ & List of resources available\\
      $dependenciesList$ & \{(t,t',d) $\mid$ $t'$ is a parent of $t$ that sends $d$ units of data\} \\
      $Pa_t$ & All the parents of task $t$\\
      $failures_t$ & Number of times task $t$ has failed to complete execution \\
      $repCount_t$ & Replica count for task $t$\\ 
      $features_t$ & Set of features for task $t$\\ 
     PCA & Principal Component Analysis\\
      COV & Coverage of Variance\\
     $\alpha_t$ &Number of completed checkpoints for task $t$\\
	$replicas_t$ & Set of replicas generated for the given task $t$\\
 	$isBusy(v)$ & Returns true if the VM $v$ is busy executing a backlog of tasks\\
\noalign{\smallskip}\hline
\end{tabular}
\end{table}

\label{sec1}

\vskip 2em
HEFT is a suitable base algorithm for scheduling the tasks of a workflow. Wieczorek et al. \cite{wieczorek2005scheduling} analyze and evaluate the performance of HEFT, Genetic Algorithms (GA) and simple "myopic" for scheduling scientific workflows. Their results show that the full-graph scheduling technique with HEFT algorithm performs best when compared to other strategies. Hence, it is decided to use HEFT to determine the initial schedule of tasks and their corresponding replicas. Furthermore, the performance of CRCH in faulty environments has been benchmarked against that of  HEFT.   
The basic HEFT algorithm does not involve any fault tolerance. If a resource fails, any task executing on that resource also fails and therefore the workflow itself fails. One approach to providing fault tolerance is to generate replicas of the tasks. Even if the task fails, the replicas can generate the intermediate results. \cite{zhang2009combined} mentions a \textit{ReplicateAll} algorithm that replicates each task by a constant factor. This redundancy improves the probability of completion of the workflow, although it increases the Total Execution Time (TET) or Makespan by a huge margin. \cite{grandstrand:2004, plankensteiner2009new} use resubmission wherein a task that could not complete execution due to a failed resource is resubmitted on another. This method suffers from the loss of computation involved in resubmitting a task that has almost completed. \cite{grandstrand:2004, zhang2009combined} use checkpointing wherein the states of the tasks are saved at regular intervals so that upon resubmission they can start execution from the previously saved state. Table \ref{tab:1} elucidates the various terms/abbreviations referenced throughout the paper.
\subsection{CRCH Algorithm}
\label{subsec1}
Figure \ref{flow_chart} highlights the data flow path, processing involved in the proposed method, CRCH. It consists of three modules namely \textit{Clustering}, \textit{Replication} and \textit{Checkpointing}. The \textit{Clustering} phase facilitates the computation of replication counts based on the properties of tasks. In the \textit{Replication} phase, the tasks are replicated by applying the replication heuristics and the standard HEFT algorithm decides the overall schedule. In the final Checkpointing and resubmission phase, a light weight strategy is used, wherein only pointers to the saved state are stored. 

\par
	 The average execution time of the task $w_t$ and the average transfer time $e(t,t',d)$ are represented using the Equations~(\ref{eq-1}) and (\ref{eq0}) respectively.

 \begin{equation}
 		w_t = \dfrac{\sum_{r \,\epsilon \, vmList}{} timeOnVm(t,r)}{vmList.size()}
 		\label{eq-1}
 	\end{equation}
 	
\begin{equation} 	
e(t,t',d) = \dfrac{\sum_{\forall \, r \, , r'  \,\in \, vmList}{}d/dataTransfer(r,r')}{|{\{(r,r')\,\,\forall \, r \, 			, r'  \,\in \, vmList}\}|}
 		\label{eq0}
 	\end{equation} 	

	Further, each task is represented in an $n$-dimensional space as a point.
	Let task $t_i$ be denoted with feature vector
	$F_i = [x_{1i}, \, x_{2i}, \, ..., \, x_{ni}]$, that is, each task has $n$ features associated with it. These features can be nominal (priority of a task) or numeric (average execution time). The axes of the $n$-dimensional space denote the various features. Some of the possible features are :

\begin{enumerate}	 					
 	\item Average execution time of a task : $w_{t}$
 	\item Average time to transfer data from the parents of the task :\[e(t) = max_{(t,t',d) \, \in \, dependenciesList} \, (e(t,t',d))\]
 	\item Priority of task 
 	\item Number of parents: \[\|\{t' \, | \,\, \exists  \, e(t,t',d) \in dependenciesList \}\|\]
 	\item Number of children: \[\|\{t' \, | \,\,  \exists \, e(t',t,d) \in dependenciesList \}\|\]				
 \end{enumerate}


\subsubsection{Clustering Module}
\label{Replication Count Algorithm}
\begin{sloppypar}
    The replication count for each task can be determined using multiple Machine Learning techniques ranging from Supervised Classification like Logistic Regression, Max Entropy Models, etc. to Unsupervised Classification like Clustering, LSA, PLSA \cite{ISLR2013, PCA2002}. In general, this can be treated as a multi-class classification problem, where the inputs are task representations and the target values are one-hot vectors $y_i$ ($y_{i_j}$=1 if replication count for the input task $X_i$ is $j$). When substantial labeled training data is present, a Multilayered Perceptron (MLP) works reasonably well. For each target class $j$, there exists a weight vector $W_j$. $P_j(t_i)$ represented by Equation~(\ref{eq1}) denotes the probability of task $t_i$ to have replication count $j$, where $F_i$ is the feature representation for task $i$.
\end{sloppypar}
    \begin{equation}   
        P_j(t_i) = \dfrac{exp(F_i \cdot W_j)}{\sum_{k}exp(F_i \cdot W_k)}
        \label{eq1}
    \end{equation}
	Equation \eqref{eq1} is a standard softmax formulation, and the MLP can be trained with a cross entropy loss (Equation \eqref{eq2}) and an optimizer like Stochastic Gradient Descent, Batch Gradient Descent, or for faster convergence RMSprop or Adam is used \cite{ISLR2013}. 
	\begin{equation}
	Loss = \frac{1}{N}\sum_{i}{}(\,-\sum_{j}{}s_{ij} \cdot log(p_{ij}))
	\label{eq2}
	\end{equation}
	where $S_i = [s_{i1},\,s_{i2},.....,\,s_{iK}]$ is a one hot encoding vector denoting the class to which the $i^{th}$ observation belongs and $P_i = [p_{i1},\,p_{i2},....,\,p_{ik}]$ is a vector denoting the probability distribution over $K$ classes predicted by the logistic classifier for the $i^{th}$ observation. 

    To generalize well, this method requires a training set of large size with already existing replication counts for a task in multiple workflows \cite{ISLR2013}. Since such a set of observations is not readily available and needs to be compiled over a period of time, a move towards an unsupervised classification scheme is adopted. One such method is \textit{Principal Component Analysis (PCA)} \cite{PCA2002} where the feature vectors are projected onto less than $K=|F|$ component vectors, so that the coverage of variance in the data is greater than a predefined threshold. PCA can be attributed to recognize correlated features, which help in representing the points in a lower dimensional space. These principal components are orthogonal unit vectors. With the addition of each such vector, the portion of variance covered with respect to the original variance in the data is improved. This improvised representation not only facilitates faster clustering but also prevents over-fitting to a certain extent \cite{PCA2002}. Since the features used are of different scales like numeric regarding the average execution time of a task or the number of parents/children of a task to ordinal in case of priority, the data needs to be standardized before the application of PCA. PCA predominantly uses correlation/ covariance matrix of the set of features. The analysis has been done on a training set of 100 points, each representing a task in a ten dimensional space, with a covariance threshold of 0.8. Intuitively, it can be seen that the number of parents is positively correlated with the average transfer time of data from parents. PCA would help remove such co-dependence. Before performing PCA, the data is mean subtracted and standardized (whitened). Since a dataset like task-features can be enormous regarding the basic characteristics of a task (execution time, priority) and its basis expansion would be even higher, a method like PCA would help identify coherent features like criticality against the number of dependents of the task. Steps 2-9 in Algorithm \ref{euclid} focus on determining the set of principal components ($K$ of them) and step 10 deals with the projection of the original points onto the new basis. 

CRCH involves hierarchical clustering on the projected points \cite{PCA2002, ISLR2013}. The distance between two clusters, characterized using the average Euclidean distance between every pair of points belonging to the two clusters is given by Equation \eqref{dist_eq}. This can be referred to as the affinity between the two clusters \cite{yang2016joint}. In each iteration, clusters are merged to form superclusters, based on their affinity to neighbors. The agglomerative clustering strategy used has been derived from the triplet loss method which has been popular in clustering high dimensional deep neural embeddings \cite{yang2016joint}.  

Triplet loss validates that the supercluster under consideration $C_m$, would be merged with one of the neighbors $C_n$ not only based on the closeness of the two superclusters but also on the additional condition that any other $C_k \neq C_n$ ($C_k$, $C_n \in \eta(C_m, R)$, where  $\eta(C_m, R)$ is the set of $R$ closest superclusters to $C_m$) is much further away from $C_m$. The pair of clusters that minimize Equation (\ref{trip_loss}) is merged into a supercluster for the subsequent time step. At any given time step, let $D_{ij}$ represent the distance between two superclusters, $C_i$ and $C_j$. Then,
\begin{equation}
    D_{ij} = \frac{
        \sum_{p_i \in C_i} \sum_{p_j \in C_j} ||p_i - p_j||_{2}
    }{
        |C_i| \dot |C_j| 
    }
    \label{dist_eq}
\end{equation}

\begin{equation}
    loss(C_i, C_j) = D_{ij}+\frac{\lambda}{R-1}\sum_{k \in \eta(C_i, R)}(D_{ij} - D_{ik})
    \label{trip_loss}
\end{equation}
The time step at which the clustering would converge is decided by the dendrogram generated by the superclusters \cite{PCA2002, ISLR2013}. At each level the branches in the dendrogram reduce, and there exists a point at which the minimum inter-cluster distance exceeds a certain threshold, the number of branches in the dendrogram at this time step is indicative of the final number of superclusters \cite{PCA2002, ISLR2013}. Once steps 11-17 are done building the superclusters, steps 18-19 deal with the assignment of replication counts, which can be decided by the size and other summary statistics of the superclusters like average execution time, average priority, etc.

\begin{figure*}[t]
    \captionsetup{justification=centering}
    \centering
    \includegraphics[width=12cm,height=8cm,keepaspectratio]{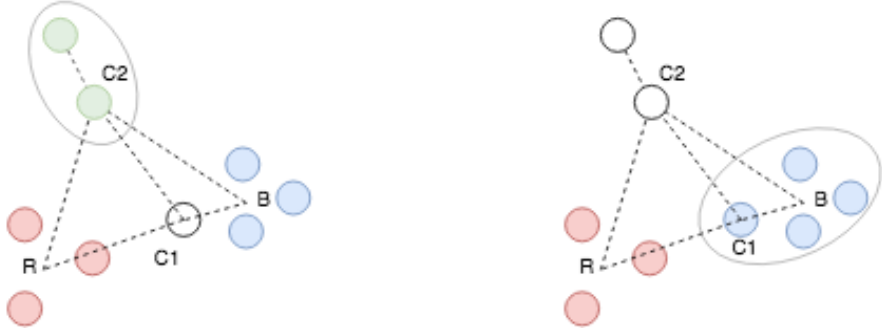}
    \caption{Action taken at a given time step by triplet loss (left) and agglomerative clustering (right). \newline (to be viewed in color)}
    \label{Triplet Loss}
\end{figure*}

\begin{figure*}[t]
    \centering
    \includegraphics[width=16cm,height=16cm,keepaspectratio]{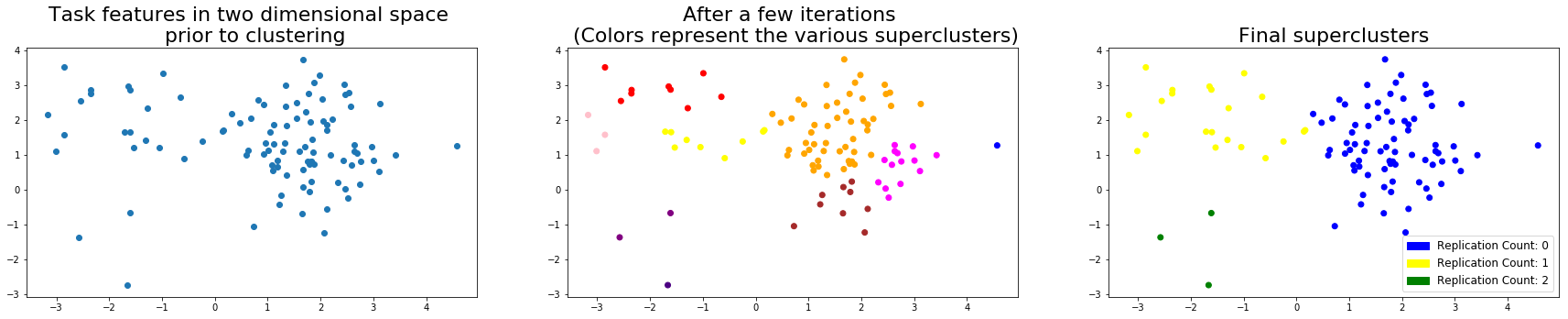}
    \caption{Clustering progress across iterations. (to be viewed in color)}
    \label{Clustering Iterations}
\end{figure*}

Figure \ref{Triplet Loss} shows the state of the agglomerative clustering procedure at a time step $t$. Each blob denotes the latent representation of a task, while the dotted lines represent the inter-cluster distances for tasks $C1$ and $C2$ (distances not impacting the loss have not been shown). Blobs in the same color belong to the same supercluster. $R$ and $B$ denote the figurative cluster centers for the superclusters in $red$ and $blue$. In a traditional clustering approach, $C1$ would end up getting merged with its closest neighbor. In the presence of triplet loss, $C2$, which is not as close as $C1$ is to its closest neighbor, but is considerably far from the remaining $R-1$ neighbors forms the green supercluster. This prevents the agglomeration from collapsing into superclusters of sizes that lie on either end of the spectrum. For example, in traditional approaches, it is not uncommon to notice agglomerative clustering from converging into either very huge clusters or very tiny ones (size$<=$3). Figure \ref{Clustering Iterations} shows the progress of supercluster formation across iterations along with the assignment of replication counts based on the final sizes of the superclusters. 

Many scientific workflows exhibit properties wherein most of the tasks can be segmented into a few large clusters. Each segment encompasses tasks having similar features, and they end up having low replication counts. The outliers based on high priority or high average execution time belong to clusters of much smaller sizes and hence are assigned a higher replication count. Such an assignment policy may lead to cases wherein a low priority task taking less time to execute (on average) gets allotted more replicas than needed as it ends up being an outlier. Simple rule ensembles, based on sufficient statistics of the feature values in the supercluster can be learned to avoid this.

\begin{algorithm*}
\begin{algorithmic}[1]

\Procedure{replicationCount(\textit{taskList, threshold, K})}{}
\State Let \textit{COV} = 0.0 
\Comment{Coverage of Variance}
\State Let \textit{principalComponents} = \{\}
\State Let \textit{dataSet} = $\{\textit{t.features} \, | \, t \epsilon \textit{taskList}\}$

\Do
    \State Let \textit{PC} = nextPrincipalComponent(\textit{dataSet})
    \Comment{Orthogonal unit vector}
    
    \State \textit{COV} = \textit{COV} + variance(\textit{PC})
    \Comment{Additional variance accounted}
    
    \State \textit{principalComponents} = \textit{principalComponents} $\bigcup$ \textit{PC} 
    
\doWhile{$\textit{COV} < \textit{threshold}$}

\State \textit{dataSet} = \textit{dataSet} $\times$ \textit{principalComponents}
\Comment{Project points in lower dimensional space}

\State Let \textit{clusters} = \textit{dataSet}

\Do

    \State Let \textit{newClusters} = hierarchicalClustering(\textit{clusters})
    \Comment{triplet loss}
     
    \State Let oldClusters = \{C $|$ C $\in$ \textit{clusters} $\wedge$  $\exists$ C' $\in$ \textit{newClusters} $\wedge$  C $\in$ C'\}

    \State \textit{clusters} = (\textit{clusters} - \textit{oldClusters}) $\bigcup$ \textit{newClusters} 

\doWhile{$|\textit{clusters}|>\textit{K}$}

\State Sort clusters based on the size of each cluster in descending order of size (or other statistics)

\For{\textit{$C_i$} $\in$ \textit{clusters}}
    
    \State $repCount_t$ = i  $\forall$ t $\in$ \textit{$C_i$} 
    \Comment{Assign replication counts}

\EndFor

\EndProcedure
\end{algorithmic}
\caption{Replication Count Algorithm\label{euclid}}
\end{algorithm*}

\subsubsection{Replication Module}
\label{Task Schedule}

The tasks part of the initial workflow graph are the original tasks, and their duplicates are referred to as replicas. Once Algorithm 1 has ascertained the replication count for each task, Algorithm 2 defines the HEFT schedule for each original task and its replicas. 

The original tasks are sorted in descending order of their estimated start times (calculated based on the inherent dependencies). Once an initial task is scheduled on a resource, the algorithm checks whether all of its original sibling tasks have also been scheduled. If so, the replicas of this initial task are mapped to resources based on which the resources would be allocated an execution interval with the minimum EST.

The following section discusses the influence of task checkpoints, on resubmissions (due to resource failures) at runtime. Here, it is assumed that there exists a subset of VMs which are entirely reliable (non-failing). Other non-reliable resources, may not be available when a task is scheduled to execute or may go down during the execution of a task. Resource failures cause delays in task execution and this results in AFT (Actual Finishing Time) being greater than EFT (Estimated Finishing Time). Such delays, when propagated along the critical path of the workflow have a direct impact on TET. Some proportions of the execution delays for tasks that are not on the critical path tends to get subsumed under the slack time that occurs between two noncritical tasks. It may also be possible that a resource is busy executing a backlog of tasks (in HEFT order), right when a task scheduled to run on that resource has met its dependencies, and as a result ends up waiting for that resource.

%
\begin{algorithm*}
\caption{HEFT with Over-Provisioning}
\begin{algorithmic}[1]
\Procedure{HEFT(\textit{taskList, dependenciesList})}{}
\For{$t \in \textit{taskList} $}
	\State Calculate $taskRank_{t}$ = B-Level of the task in critical path calculation	
\EndFor

\State sort $taskList$ based on $taskRank_{t}$ $\forall \,t \, \epsilon \, taskList $ in descending order

\For{$t \in \textit{taskList} $}
	\State Schedule $t$ on $v$ with minimum $EST_t$
	\If{$(t,t',d) \in \textit{dependenciesList}$}
		\If{$t'' \, is \, scheduled \, \, \forall \, (t'',t',d) \in \textit{dependenciesList}$}
			\State Schedule replicas of $t'$ on VMs with minimum ESTs
		\EndIf		
	\EndIf
\EndFor

\EndProcedure
\end{algorithmic}
 \label{heftOP}
\end{algorithm*}

\subsubsection{Checkpointing Module}

Algorithm \ref{checkpointHEFT} defines the execution of a workflow at runtime concerning checkpointing and resubmission.
The environment is modeled to have a set of resources which fail depending on their respective Mean Time To Repair (MTTR) and  Mean Time Between Failures (MTBF). The size and duration of these failures is dependent on the type of environment (\textit{Stable}, \textit{Normal} or \textit{Unstable}). In the case of a perfect environment with no failures and assuming no checkpoints are involved, the TET ($TET_{perfect}$) is determined by the task schedule, as described in section \ref{Task Schedule}. At any given point, let $\alpha_i$ be the number of checkpoints that have been completed for $t_i$. If task $t_i$ is scheduled to start on $v_j$ at $EST_i$ and has $runtime_{ij}$ then, $TET_{perfect}$ is represented using Equation~\ref{TET_perfect}.
\medmuskip=0mu
\begin{equation}
  TET_{perfect} = \max_{t_i \,on \,v_j}(EST_i + \,runtime_{ij})\,\forall \,\,i \in taskList
  \label{TET_perfect}
\end{equation}
A schedule S is a set of tasks with $t_i$ scheduled on a VM $v_i$ from $EST_i$ to $EFT_i$. Only when $v_i$ belongs to a set of Failing VMs ($FVM$), $t_i$ may or may not be rescheduled on a nonfailing VM, that is, VM $\notin$ FVM. The two cases to consider here are whether $v_i$ fails during the execution of $t_i$ (Case 1) or $v_i$ is down at $EST_i$ (Case 2). $\alpha_i$ is the number of checkpoints that have been completed for $t_i$. These checkpoints are global and are synchronized, that is, at each checkpoint, the working memory is dumped to a nonvolatile stable storage associated with each VM. When a task $t_i$ completes its execution on $v_i$, the result of its execution is stored in the stable storage associated with $v_i$. The pointer to the location on stable storage is stored in a global memory. This pointer can be referenced using a hash value of the task id for quick access. Along with the pointer, if $v_i$ $\in$ $FVM$, the results are also stored in the global memory. If $children_i$ is the set of tasks that use the results generated from $t_i$, then the data can be fetched by $child_k$ $\in$ $children_i$ using pointers obtained from dereferencing the global memory. In case the VM on which the parent was executed is down, the data can be fetched directly from the global memory. Each VM $v$ $\in$ $FVM$ has a set of time intervals denoted by $L_v$ which indicate the intervals in time when the resource would be unavailable to execute a task that has been scheduled on it (using $MTTR_v$, $MTBF_v$ sampled from log-normal and Weibull distributions respectively).

A task may fail to comply with the HEFT schedule for one of the following reasons.\begin{itemize}
    \item Resource fails during the execution of a task.
    \item Task is scheduled to start on a VM that is currently down.
    \item Task is scheduled to start on a VM that is executing a backlog of tasks.
\end{itemize}

\begin{algorithm*}
\caption{CheckpointHEFT} 
\begin{algorithmic}[1]
\While{$\exists$ \textit{t} such that $t$ is not scheduled}
 \State schedule \textit{t} on VM \textit{v} using Algorithm \ref{euclid}, Algorithm \ref{heftOP}
 \If {isBusy($v$)}
	\If {$ \displaystyle EFT_t = max_{t' \in replicas_t} EFT_{t'}$} 
	\Comment $t$ is the last unscheduled replica%
		\State Wait for $v$ to be online	
	\EndIf
 	\Else
 		\State Terminate $t$
        \State $failures_t \gets failures_t + 1$
 \EndIf
  \If {$v \in FVM$}
    \If {\textit{t} has started execution on \textit{v}}
      \State $(X,Y) \gets \argmin_{(x,y)  \in L_v  \, \wedge  \, x  \geq  AST_t} (x - AST_t)$
       \If {$AFT_t \leq X$} \Comment The task finishes before X 
        \State Continue
       \EndIf
      \State $failures_t \gets failures_t + 1$
      \If {$failures_t$ = $repCount_t$}
        \State $minEST \gets minimum \; EST\;  for\;  t\;  on\;  a\; VM \; v_{new} \; st\; v_{new} \;is \; non failing $
        \State $vmWithMinEST \gets v_{new}$
        \State $executionSaved \gets t.\alpha \times \lambda$
        \State $overhead \gets executionSaved$
          \If {($minEST$ + $overhead$) $<$ Y}
            \State $Schedule(t,vmWithMinEST)$
          \Else
            \State $Schedule(adjustExecutionTime(t),v)$
            \Comment{Reduced Execution Time}
          \EndIf
      \EndIf
 
 \Else
    \Comment $t$ is scheduled on a resource that is currently down %
    \State $failures_t \gets failures_t + 1$
    \If {$failures_t$ = $repCount_t$}
        \State $(X,Y) \gets \argmin_{(x,y) \in L_v  \, \wedge  \, x\leq AST_t} (AST_t-x)$
        
        \State $minEST \gets minimum \; EST\;  for\;  t\;  on\;  a\; VM \; v_{new} \; st\; v_{new} \;is \; non failing $
        \State $vmWithMinEST \gets v_{new}$
        
        \If {$minEST$ $<$ Y} 
          \State $Schedule (t,vmWithMinEST)$
        \Else
          \State $Schedule(t,v)$
      \EndIf
    \EndIf  
\EndIf 
\EndIf
\EndWhile
\end{algorithmic}
\label{checkpointHEFT}
\end{algorithm*}

 \textit{Light-weight checkpointing} reduces the load on the global memory and thus reduces the chance of a possible bottleneck that can result from a large number of memory accesses. 
 
 Algorithm \ref{checkpointHEFT} takes as input the HEFT schedule defined by Algorithm \ref{euclid} and Algorithm \ref{heftOP}. There is a possibility that a task $t_i$ scheduled on a failing VM $v_i$ at $EST_i$ cannot begin execution because of a backlog of tasks that are currently executing on $v_i$. If $t_i$ is the last replica of a given task which has not executed successfully, it needs to wait until a VM is available, else it can be terminated and deemed as a failure (Steps 3-8). Tasks that have their dependencies satisfied and yet can't be executed on their scheduled resource due to the backlog of tasks being executed on them can be deemed as failures. In unstable environments where a decent proportion of tasks aren't able to find their scheduled resource free, even when their dependencies are fulfilled, it has been found that it is better not to consider them as failures, as it would lead to unnecessary resubmissions on a set of few non failing VMs, thus forming a bottleneck. This would simply increase resubmissions and would not exploit the usage of replicas. Step 9 checks if a task is scheduled on an unreliable VM, and if so, whether the task has begun execution or not. Step 11 defines variables \textit{X,Y} which represent the points in time during which the resource in question is down. If the task completes before the next time interval when the resource is down, execution proceeds normally (Steps 12-13). Steps 16-17 identify the nonfailing VM with the minimum EST whereas steps 18-19 calculate the proportion of the task completed (in terms of execution time),  which would constitute the overhead involved in scheduling the task on a different resource. If the task is to be scheduled on the same VM it can resume from the latest checkpoint, if not it pays the overhead of re-executing already completed checkpoints. Steps 20-23 describe this comparison. Steps 14-15 and 25-26 ensure that a task gets resubmitted only if and only if all its preceding replicas including itself have failed execution.

The case when the resource is down at $EST_i$ for $t_i$ is dealt with by comparing the minimum possible EST at a non-failing VM with the point in time when the currently down resource would be available at the latest (Steps 26-32). 

Since failure count of an original workflow task is a task level attribute, multiple processes running on different VMs may contest to update the count simultaneously. This justifies the need for a cache coherence strategy like MESI to avoid inconsistencies.

\subsection{Dynamic Checkpoint Interval}
\label{Dynamic CP}
\begin{sloppypar}
The checkpoint interval $\lambda$ is a global parameter which defines the interval between two consecutive checkpoints. Each checkpoint has an overhead $\gamma$ associated with it. To improve the TET, $\lambda$ can hold an optimal value depending on the environment of execution. In a highly stable environment failures are rare. Hence TET would reduce with a higher $\lambda$. Hence higher $\lambda$ would mean less number of total checkpoints and reduced total checkpoint overhead. In the case of an \textit{unstable} environment with a large number of failures, TET may improve with a smaller $\lambda$, as the data lost from having checkpoints further apart would be the delimiting factor instead of the overhead of checkpoints.
\end{sloppypar}

Checkpoint interval analysis is heavily dependent on the critical path. A critical path of a DAG is a path from an entry node to an exit node, whose length is the maximum \cite{kwok1999benchmarking}. To compute the critical path of the workflow after the addition of replicas and for determining the schedule using Algorithm \ref{heftOP}, a simple backtracking approach is sufficient. Let $V_{max}$ be the VM on which the task $t_{max} = \argmax_t {EFT_t}$ is executed. Since $t_{max}$ is scheduled on a VM with minimum possible $EST_{t_{max}}$, there exists at least one such task $t_{max-1}$, where $EFT_{t_{max-1}}=EST_{t_{max}}$. By induction, the critical path can be backtracked.

This section highlights a proof which suggests a varying checkpoint interval for an optimal TET with the same level of fault tolerance. It assumes that there is no shift in the critical path with the involvement of resubmissions.

\begin{lemma}
\label{lemma_1}
The checkpoint interval $\lambda$ introduced in the CRCH algorithm is environment dependent and can be modified to best suit the TET of the workflow in a given environment. 
\end{lemma}
\begin{assumption}
\label{as1}
The waiting time of a task is distributed normally, and the probability density function depends on the ancestors of the task, environmental parameters. The range of values observed can only be explained by a normal distribution which spans $[\mu-3\sigma,\mu+3\sigma]$. The set of observations for waiting times of tasks taken over a large range of inputs closely mimicked a normal distribution, indicated by a QQ (quantile-quantile) plot \cite{natrella2010nist}. Upon reasonable \& realistic variations in $\lambda$, while maintaining the same environmental conditions it was found that the shift in the distribution parameters is negligible.
\end{assumption}
\begin{assumption}
\label{as2}
The probability of a task to fail on the scheduled resource is independent of $\lambda$.
\end{assumption}
\begin{assumption}
\label{as3}
The resubmission of ancestor tasks does not influence the probability of a task to fail on the resource it is meant to execute on. 
\end{assumption}

We go on to prove Lemma \ref{lemma_1} based on the aforementioned assumptions. Notations used in the proof are as follows :

\begin{itemize}[]
     \item[] $\lambda$ : checkpoint interval (time interval between two successive globally synchronized checkpoints )
     \item[] $\gamma$ : overhead at each checkpoint
     \item[] $TET_{CRCH}$ : expected time to execute CRCH algorithm (equation \ref{tet_crch}) 
     \item[] $CO$ : checkpoint overhead (equation \ref{co}) 
     \item[] $TET_{CRCH/CO}$ :  expected time to execute CRCH excluding $CO$ (equation \ref{tet_crch_co})
     \item[] $TET_{C_{i}}$ : expected time taken for task $t_{i}$ to execute in CRCH (eqution \ref{tet_ci})
     \item[] $CP$ : critical path - a set of tasks lying on the critical path
     \item[] $WT_{i}$ : expected waiting time for task $t_{i}$
     \item[] $RO_{i}$ : expected resubmission time for task $t_{i}$
     \item[] $TET_{H_{i}}$ : time taken for task $t_{i}$ to execute in HEFT
\end{itemize}

\begin{proof}
According to the definitions mentioned above:
\begin{equation}
 TET_{C_{i}} = TET_{H_{i}} + WT_{i} + RO_{i} \label{tet_ci} \\
\end{equation}
\begin{equation}
   TET_{CRCH/CO} = \sum_{i \in CP} TET_{Ci} \label{tet_crch_co}\\
\end{equation}
\begin{equation}
     CO = TET_{CRCH/CO} \times \dfrac{\gamma}{\lambda}\label{co}\\
\end{equation}
 \begin{equation}
 TET_{CRCH} = TET_{CRCH/CO} +  CO \label{tet_crch}
\end{equation}
 If the task $t_{i}$ is the last of its replicas, it may not follow the schedule given by Algorithm \ref{heftOP} due to:
 
\begin{enumerate}
    \item All the parents in $Pa_i$ haven't completed execution that is, there exists one task
    $t_p$ $\in$ $Pa_i$ st. $\forall$$t_j$ 
    $\in$ $replicas(t_p)$, $t_j$ hasn't completed execution.
    \item The resource on which the task has been scheduled is executing some backlog of tasks.
    \item The resource is unavailable, as it has failed.
\end{enumerate}

 $WT_{i}$ (equation \ref{wti}) depends on $A(i)$, that is, the ancestor graph for $t_{i}$ (which includes the VMs on which the tasks have been scheduled \& their corresponding MTBFs, MTTRs). $ N_w(\mu_{w},\sigma_w)$, is a Gaussian distribution which determines the expected amount of time a task has to wait on its immediate parents and the sufficient statistics are functions of $A(i)$. (Assumption \ref{as1}) Thus:
\begin{equation}
 WT_i \sim N_w(\mu_{w}(A(i)), \sigma_{w}(A(i)))
  \label{wti}
\end{equation}
 $RO_i$ is the overhead involved in resubmission. A task $t_i$ is resubmitted only if it is the last replica and all the other replicas have failed, that is,
 \begin{align}
  EFT_{i} = \max_{ r \in replicas(i)} EFT_{r}
  \label{eft_i}
  \end{align}
The following calculations for the probability of failure of a task and the corresponding expected resubmission overhead are only done for tasks compliant with equation \eqref{eft_i}. 

The probability that a task $t_i$ fails, defined as $P_{t_i}$, depends on the probability that the task has been scheduled on a failing VM $v_i$ and the probability for an intersection of the execution time with the interval $L_{i}$ where
\begin{align}
    L_{i} = \{(X,Y): Y-X = mttr_{i} \text{ and $v_i$ fails during $(X,Y)$}\}
\end{align}
Since the set of failing VMs is chosen from uniform distribution:
\begin{equation}
    P_{v_i} = \dfrac{|FVM|}{|V|}  \label{6}
\end{equation}
\begin{description}
\item $P_{v_i}$ is the probability of a VM $v_i$ to fail
\item $FVM$ is the set of failing VMs
\end{description}

 Now to evaluate the probability of interval overlap, $MTBF_{i}$ and $MTTR_{i}$ are needed which are determined from Weibull distributions $W_{MTBF}$ and $W_{MTTR}$. Thus:
\begin{equation}
    P(overlap) = g(EST_i,EFT_i,W_{MTBF},W_{MTTR}) \label{7}\\
\end{equation}
\begin{equation}
     Using \, \eqref{6}  \eqref{7} : \,\, P_{t_i} = P(overlap)\cdot \dfrac{|FVM|}{|V|} \label{8}
\end{equation}
As the executions of all replicas are mutually independent, with $NF_i$ being the number of failing replicas of $t_i$ :
\begin{equation}
    P(NF_i = R_i) = \prod_{
    \begin{small}
        r \in replicas(i)
    \end{small}} P_{t_r} \label{9}
\end{equation}
    \begin{description}
    \item $NF_i$ is the number of failing replicas of task $t_i$
    \item $R_i$ is the total number of replicas of $t_i$
    \end{description}
If $E(RO)$ is the expected overhead in resubmissions then by using \eqref{9}:
\begin{equation}
    RO_i = P(NF_i = R_i) \cdot E(RO) \label{10}
\end{equation}
 Let $new$ be the resource on which $t_i$ is resubmitted. Hence $P(new  = v_i)$ is the probability that the task is resubmitted on the same resource $v_i$ on which it was originally submitted. $new$ follows a multinomial distribution conditional on the schedule decided by \ref{heftOP} and any other workflow or resource parameters (including $\lambda$). Overhead due to submitting on the same resource:
\begin{equation}
    Overhead_{same} = E(minEST_{same}) + PF_i - \bigg\lfloor\dfrac{PF_i}{\lambda}\bigg\rfloor\cdot \lambda \label{11}
\end{equation}
\begin {description}
\item where
\item $PF_i$ is the point in time where the execution of $t_i$ was stopped on $v_i$ due to failure with respect to $EST_i$.
\item $E(minEST_{same})$ is the expected value for the difference in minimum Estimated Start Time when the task is rescheduled on the same resource and $PF_i$ (point of failure).
\item $E(minEST_{diff})$ is the expected value for the difference in minimum Estimated Start Time when the task is rescheduled on a different non-failing resource and $PF_i$ (point of failure).
\end{description}
 Therefore, the probability of the task $t_i$ being resubmitted on a different resource is $ (1 - P(new  = v_i)).$ Overhead due to submission on a different resource:
\begin{equation}
    Overhead_{diff} = E(minEST_{diff})+TET_{Hi} \label{12}
\end{equation}
Using \eqref{9}, \eqref{10}, \eqref{11}, \eqref{12}:
\begin{equation}
    \begin{split}
        RO_i = P_{t_i}^{R_i}\cdot[P(new = v_i)\cdot Overhead_{same} + (1-P(new=v_i))\cdot Overhead_{different}]
        \end{split} 
    \label{13}
\end{equation}
That is, 
\begin{small}
\begin{align}
    \begin{split}
    RO_i = P_{t_i}^{R_i}\cdot\bigg[P(new = v_i)\cdot \big\{ E(minEST_{same}) + PF_i - \bigg\lfloor\dfrac{PF_i}{\lambda}\bigg\rfloor\cdot \lambda\big\} 
        &\\+ (1-P(new=v_i))\cdot \big\{E(minEST_{diff})+TET_{Hi}\big\}\bigg]
    \end{split}
    \label{14}
\end{align}
\end{small}
Using \eqref{tet_ci}, \eqref{tet_crch_co}, \eqref{co}, \eqref{tet_crch}, \eqref{14}
\begin{small}
\begin{multline}
     TET_{CRCH/CO} = \sum_{i \in CP}\bigg(TET_{Hi} + \mu_w(A(i))
        \\+P_{t_i}^{R_i}\cdot\bigg[P(new = v_i)\cdot \big\{ E(minEST_{same}) + PF_i - \bigg\lfloor\dfrac{PF_i}{\lambda}\bigg\rfloor\cdot \lambda\big\} 
        \\+ (1-P(new=v_i))\cdot \big\{E(minEST_{diff})+TET_{Hi}\big\}\bigg]\bigg)  
        \label{15}
\end{multline}
\end{small}
From \eqref{co}, \eqref{tet_crch}
\begin{equation}
    TET_{CRCH} = TET_{CRCH/CO}\cdot\bigg[1+\dfrac{\gamma}{\lambda}\bigg]
    \label{16}
\end{equation}
It can be seen that $TET_{CRCH}$ is a product of two separate terms involving $\lambda$. Let $Term 1 \leftarrow TET_{CRCH/CO}$ and $Term 2 \leftarrow \bigg[1+\dfrac{\gamma}{\lambda}\bigg].$ For $Term 2$, it is clear that as $\lambda$ increases the value of $Term 2 $ decreases. In $Term 1$ the expressions involving $\lambda$ are $ PF_i - \bigg\lfloor\dfrac{PF_i}{\lambda}\bigg\rfloor\cdot \lambda$, $P(new = v_i)$, $E(minEST_{diff})$, $E(minEST_{same})$ where $PF_i$ is independent of  $\lambda$ (Assumptions \ref{as2} and \ref{as3}). The first expression among these is a non differentiable function in $\lambda$, although it is piece-wise continuous. The average value of the function over a contiguous range of values for $\lambda$ (where the function is continuous and $\lambda$ is chosen randomly from a uniform distribution) would increase as $\lambda$ increases.

 In \textit{stable} environments, $P_{t_i}^{R_{i}} \ll 1.$ Therefore $Term1$ is insignificant and $\lambda$ can be increased. This means a greater checkpoint interval is affordable in a \textit{stable} environment. On the contrary, in an \textit{unstable} environment $P_{t_i}^{R_{i}}$ is considerable and cannot be ignored. $P(new = v_i)$ decreases as $\lambda$ increases since the overhead involved in re-scheduling on a different resource would be comparable to the loss in execution time units incurred from distant checkpoints. In general, with low resource availability there are many instances of resubmissions, which leads to an increase in $E(minEST_{diff})$ and $E(minEST_{same})$. Thus a lower value of $\lambda$ would be needed to reduce the overhead caused due to checkpointing.
 
 For cases where $\lambda \ll 1$, $WT_i$ would reduce as the resubmission would not account for much loss in execution time units. In $unstable$ environments, the contributing factor to the waiting time of a task would be the unavailability of resources to schedule ancestor tasks. Hence the observations on waiting times for tasks indicated a negligible impact on changes in $\lambda$. For $stable$ environments the number of resubmissions are low. Hence the waiting time of a task is not significantly influenced by changes in the parameter $\lambda$. (Assumption \ref{as1})

\end{proof}

\section{Performance Analysis}

\label{performance_analysis}

\subsection{Experimental Setup}
To evaluate the performance of the proposed approach, the WorkflowSim simulator is used. The cloud environment is modeled with one data center and 20 Condor VMs. It is assumed that there are at least four reliable VMs at any point in time.
Each VM is assumed to be fully connected to every other VM using two-way dedicated lines. This assumption is based on the fact that if they are connected using a single interconnect bus then wait time to access and gain control over the bus also needs to be considered and becomes a substantial factor in TET as compared to factors of significance such as transmission time, checkpoint overhead etc.

 Four different workflows: Montage, Cybershake, Inspiral and SIPHT are used in the simulation, with sizes ranging from 100 to 700, in multiples of 100. The workflows are given as input in the form of a DAX file. The dependencies between the tasks in the workflow, runtime of the tasks on multiple VMs,
the rate of file transmission between VMs and the size of the files to be transmitted between the tasks are read from the DAX file. Each DAX file is executed ten times and the average of the results in those executions is considered.

In some cases, the same DAX file shows high run times on a few executions and low on the others. This is attributed to the fact that if a critical task gets scheduled on a failing VM, the runtime increases by a huge margin. A task is a critical task if that lies on the critical path and has zero free and total float \cite{ISLR2013}. An increase in the runtime of this task would also increase the final execution time linearly. 

The runtimes of the tasks are different on each of the Condor VMs. Each task has a set of defined parent and children tasks and the data to be transferred between tasks is specific and predefined. Thus three different matrices are used as input. 
\begin{enumerate}
    \item $(Task \times Task)$: Data to be transferred.
    \item $(Task \times VM )$: Run time of a task on a given VM.
    \item $(VM\times VM)$: Transmission time between two VMs. 
\end{enumerate}

 A light-weight checkpointing \cite{duan2005dee} model is proposed. Checkpoints of a task only contain the program state and pointers to a data store (not the input data itself). 
Each VM has a non-volatile stable storage. Each globally synchronized checkpoint dumps the working memory of the VM to its non-volatile storage. It involves the instruction pointer, register state, current stack etc. There also exists a global storage which holds references to the outputs generated by the parents of a task. The parents’ output is stored on the resource, on which the parent was executed. Hence the local checkpoint need only save the light-weight program state. The references pointers to the locations where an output of a parent task is stored can be retrieved from the global reliable storage. The data required for execution is fetched from these references as and when required. Instead of fetching all the data from its references before the execution of a task, wastage is reduced in the transmission of unwanted data by fetching only those pages which are required at any given point during execution. Since dedicated lines exist between the VMs, the time to fetch data from disk belonging to another VM is comparable to fetching data from its own disk.

 The resource failures for a simulated environment consider Mean Time Between Failures (MTBF), the size of the failure in terms of the number of VMs affected, and the Mean Time to Repair (MTTR). Three different failure models \textit{Unstable, Normal \& Stable} have been modeled by drawing values from the following distributions. For each environment, the hyper-parameters of the distribution vary. The MTBF is modeled using a Weibull distribution with shape parameters ranging from 11.5 to 12.5 \cite{plankensteiner2009new}.
The size of the failure is also modeled using a Weibull distribution with shape parameters ranging from 1.5 to 2.4 \cite{plankensteiner2009new}. The set of failing VMs and MTTR are modeled using uniform and log-normal distributions respectively \cite{plankensteiner2009new}. The shape parameters used for MTTR
are 10 and 5.  MTBF values decrease as one moves from \textit{stable} to \textit{unstable} environment as failures are more frequent. The MTTR values (in minutes) are roughly 6/3/1 for \textit{unstable}/\textit{normal}/\textit{stable} environments. 

\subsection{Experimental Results}
\label{experimental_results}
The performance of CRCH is compared with HEFT and another heuristic approach called \textit{ReplicateAll} which was  proposed in \cite{plankensteiner2007fault}. \textit{ReplicateAll} uses a simple approach for task replication wherein each task in the workflow is replicated \textit{repCount} number of times. \textit{repCount} is assumed to be three for \textit{ReplicateAll}, and hence it is referred to as \textit{ReplicateAll(3)} in this work. Both are run on multiple workflows with the same failure models as that of CRCH. A majority of the executions for HEFT could not complete due to the absence of fault tolerance. Average Total Execution Time (TET), Resource Usage, Resource wastage and the Standard Length Ratio (SLR) are the metrics used for evaluating the performance of CRCH. 

A resource is considered good if a majority of the tasks have a very low execution time on it as compared to other VMs. If the VM chosen to fail is good, lots of tasks are likely to be scheduled on it (especially in a sparse workflow) and hence more resubmissions. Otherwise, the number of tasks originally scheduled on that resource is itself low, hence lesser number of resubmissions and a comparatively smaller increase in TET. The goodness of a failing VM and consequentially the number of resubmissions are a decisive factor in all the metrics mentioned and thus explain a majority of the anomalies.

\par
\begin{figure*}
\centering
\includegraphics[width=0.75\textwidth]{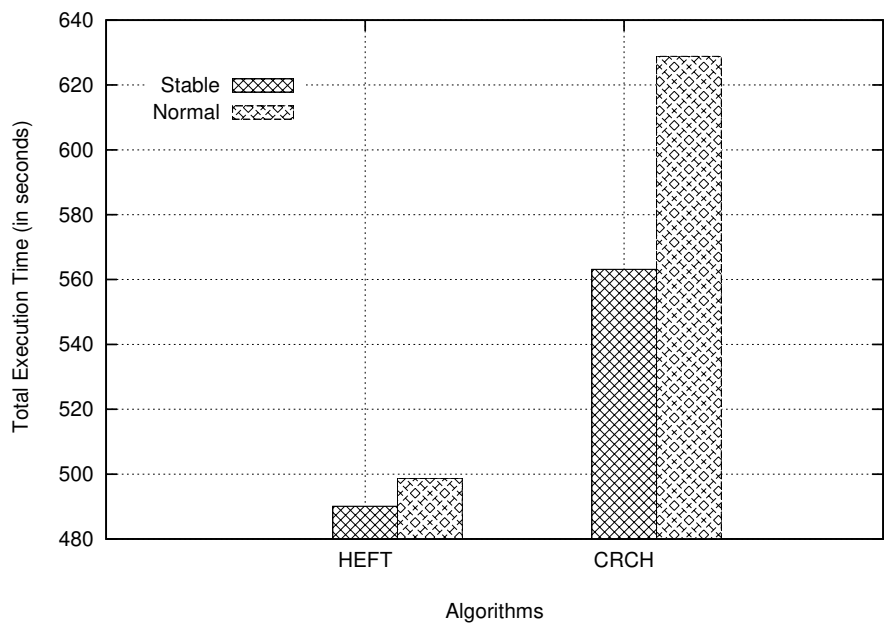}
\caption {Total Execution Time}
\label{TET}
\end{figure*}

 The results for Montage workflow have been presented but similar patterns were observed for SIPHT, LIGO and  CyberShake. TET of CRCH is greater than HEFT in all the environments. This is expected as CRCH involves execution of replicated tasks and at the time of resource failures, CRCH waits for the failed resource to be back online or schedules the task on a less optimal but available resource. This increase is more profound in the \textit{normal} environment, as the rate of resource failures is higher than that of the \textit{stable} environment. The TET of \textit{ReplicateAll(3)} in all
three environments (not shown) is much higher than its counterparts, as all the tasks of the workflow have to be executed four times. Albeit, the replicas can be scheduled in parallel, \cite{zhang2009combined} suggests a task schedule where replicas for a task are scheduled after its children. Moreover, a large number of tasks on the same level (as seen in many scientific workflows) \cite{deelman2015pegasus} would generate more replicas than that can be handled by existing VMs. These factors lead to a manifold increase in TET for \textit{ReplicateAll(3)}.
Figure \ref{TET} presents the comparison of HEFT and CRCH algorithms under \textit{stable} and \textit{normal} environments. HEFT fails to execute in the \textit{unstable} environment because of a higher probability of resource failure and lack of fault tolerance. TET in \textit{stable} is, in general, lower as compared to the \textit{normal} environment as there exist very few cases of resubmission, whereas replication amounts to the same time in both the environments. In general, as CRCH considers resubmission on the same resource, TET values are considerably lower than what can be expected, if the resubmissions were always on a different resource.

\par
\begin{figure*}[t]
    \centering
    \includegraphics[width=1.0\textwidth]{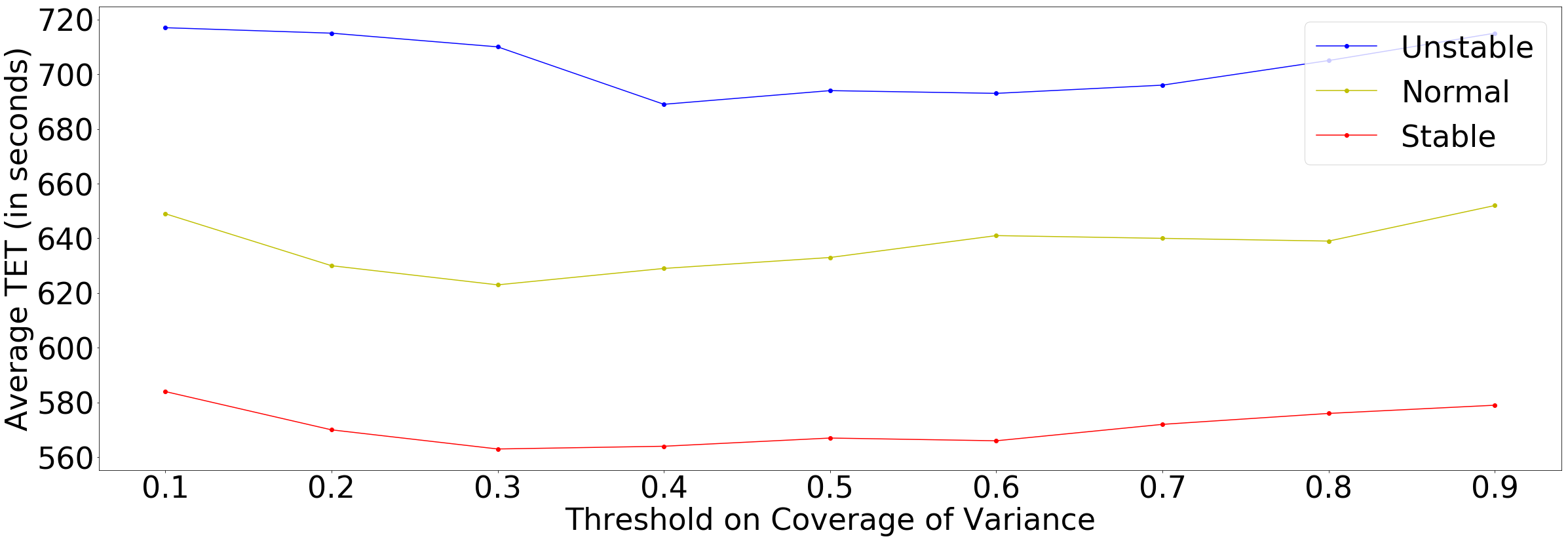}
    \caption{Clustering Overhead [Coverage of Variance]}
    \label{TET_cl_overhead}
\end{figure*}

\begin{figure*}[t]
    \centering
    \includegraphics[width=16cm,keepaspectratio]{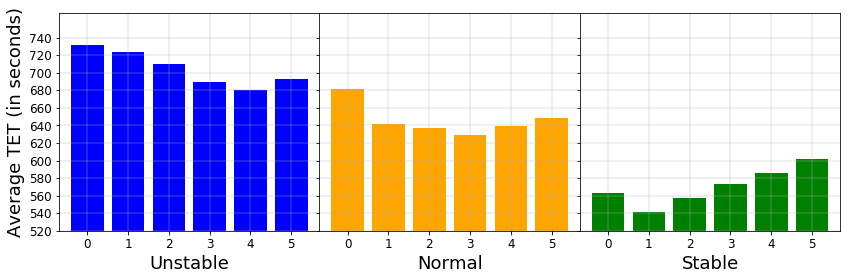}
    \caption{Clustering Overhead [Max. Replication Count]}
    \label{TET_cl_overhead_2}
\end{figure*}

 Figures \ref{TET_cl_overhead} and \ref{TET_cl_overhead_2} analyze the impact of the parameters involved in the \textit{Clustering Algorithm} on the average TET of a workflow (at optimum values of checkpointing parameters). Prior to hierarchical clustering Algorithm 1 transforms the task features to a lower dimensional space using PCA. The stopping criterion for PCA is determined by a threshold on Coverage of Variance (COV). We hypothesize that task features aren't completely independent and are prone to noise. Thus using all the features to estimate task affinities would lead to suboptimal cluster assignments. This is confirmed by an increase in average TET as we increase the threshold on COV. Similarly using lower number of features would lead to information loss. We observe that a threshold of $0.3-0.4$ leads to cluster assignments with more accurate estimates of the replication counts. The other parameter concerning clustering is the max. replication count or the number of final super clusters. An increase in this hyper-parameter leads to an expected increase in TET across all environments. It is interesting to note that the average TET doesn't increase dramatically when one moves away from the optimum max. replication count as the new cluster that is formed has a relatively small size. Having zero or very few replicas also leads to an increase in TET which can be attributed to the resubmission overhead.

\par
\begin{figure*}[t]
    \begin{subfigure}{.45\textwidth}
        \centering
        \includegraphics[width=\textwidth,keepaspectratio]{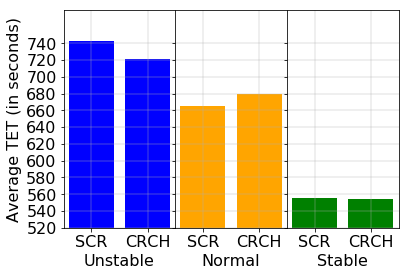}
        \caption{Comparison with SCR}
        \label{scr_comparison}
    \end{subfigure}%
    \begin{subfigure}{.45\textwidth}
        \centering
        \includegraphics[width=\textwidth,keepaspectratio]{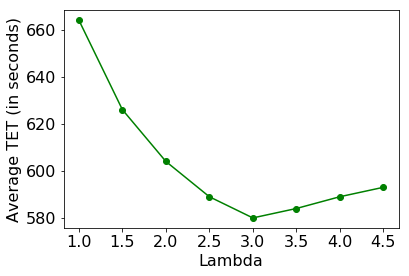}
        \caption{Impact of Checkpoint Interval}
        \label{lambda_comparison}
    \end{subfigure}
    \caption{Checkpoint Overhead}
\end{figure*}

 Figure \ref{scr_comparison} compares the performance of the Checkpoint/Restart algorithm (Algorithm 3) against the SCR Algorithm when the former involves no replicas for any task. In an unstable environment CRCH's checkpointing marginally outperforms SCR while the checkpointing overhead is comparable in the normal and stable environments. In the case of SCR, due to multiple node failures in an unstable environment the workflow needs to rely on the infrequent stable checkpoints which save the intermediate states on a Parallel File System. Restarting tasks from these points leads to excessive resource wastage and the workflow takes longer to complete. On the other hand, CRCH can afford to have more frequent light-weight checkpoints (smaller $\lambda$). In most workflows, a child task requires the data computed by its parent to restart from a saved program state. Since this data is always dumped to a stable storage upon the completion of a task and can also be accessed via a pointer saved on a stable global storage, the recovery is less expensive. For a constant $\lambda$ across all environments SCR outperforms CRCH under stable conditions, primarily due to overlapping computations and data transfers. But since $\lambda$ has been proposed to be a function of the failure parameters (MTTR, MTBF) an increased $\lambda$ in a stable environment leads to a TET that is comparable with SCR. Figure \ref{lambda_comparison} shows that the average TET is quite sensitive to $\lambda$ in a stable environment when no task has any replicas. A smaller $\lambda$ leads to frequent checkpoints while a larger value leads to increased re-computations.

\par
\begin{figure*}[!ht]
 \centering
\includegraphics[width=0.75\textwidth]{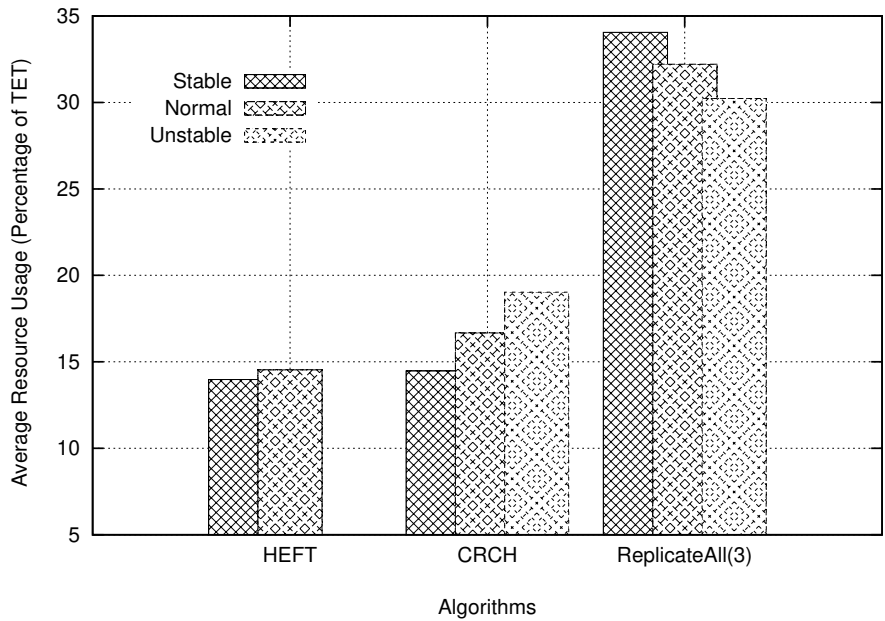}
    \caption{Average Resource Usage}
    \label{Average Resource Usage}
\end{figure*}

 Resource Usage is the sum of the actual processor time in minutes used for executing all the tasks of the workflow \cite{zhang2009combined}. Figure \ref{Average Resource Usage}  shows the average resource usage of the CRCH, HEFT and
\textit{ReplicateAll(3)} algorithms as a fraction of TET. HEFT schedules the task of the workflow in the most optimum manner \cite{topcuoglu2002performance} and thus has a minimum resource usage. HEFT cannot finish the execution of workflows in the \textit{unstable} environment, as it has no fault tolerance. CRCH, on the
other hand takes into account resource failures and uses replication and resubmission of tasks, and thus the resource usage increases. In case of the \textit{stable} environment, the resource usage
of CRCH is higher than HEFT by 16\%. This small increase is attributed only to the replication of tasks, as resubmission is occasionally done due to the low resource failure rate. However, in the case of a \textit{normal} environment, the increase is at 17\%, which is
much more prominent as along with replication, resubmission is also considered. Resubmission of the tasks increases as the failure rate of the resources increases. All the tasks are replicated
three times in \textit{ReplicateAll(3)}. That leads to higher resource usage. In the \textit{stable}, \textit{normal} and \textit{unstable} environments the average resource usage increase compared to CRCH is 41\%, 30\% and 17\% respectively. This is attributed to the fact that CRCH uses Replication Heuristics in the form of hierarchical clustering. A decline in the percentage increase in resource usage for \textit{ReplicateAll(3)} over CRCH is observed across the environments. This observation is due to the fact that when a large number of resources are failing most of the replicated tasks do not complete execution. Hence the futile usage of resources to re-execute completed tasks reduces. Instead the resources are idle. The increase in usage is highly dependent on the workflow types.

\par
Two major types of resource wastage are considered in CRCH:
\begin{enumerate}
    \item When a task is running on a resource and the resource fails,  wastage occurs when the task has executed beyond a certain checkpoint and needs to be executed on a different resource. The processing power wasted in executing a task beyond the last checkpoint is the effective wastage.
    \item When a task is replicated and the first replica executes successfully, any subsequent replica executed on any of the resources are now deemed as waste.
\end{enumerate}
\sloppy In case of HEFT where there is no resubmission, there is a possibility of a workflow failure - all tasks executed as part of the workflow are considered as waste. The wastage in HEFT is calculated over ten executions, where all failed workflow executions contribute to the waste and the wastage is averaged over the executions. There is no wastage in an execution that is completed. 
No replication heuristic is considered in \textit{ReplicateAll(3)} for deciding the number of replicas of tasks and so a lot of unnecessary replicas are created. Resource wastage in \textit{ReplicateAll(3)} is the time that is spent for executing the unnecessary replicas of tasks, after their first
replicas have already executed successfully. The resource wastage that occurs due to resubmission is eliminated, as there is no resubmission of tasks in \textit{ReplicateAll(3)}.
The average resource wastage of the HEFT, CRCH and \textit{ReplicateAll(3)} as a fraction of TET is compared in Figure \ref{TET}.
On average, CRCH gives a 22\% and 46\% reduction
in the resource waste in the \textit{normal} and \textit{stable} environments over HEFT respectively.
The improvement in resource wastage over HEFT in the \textit{normal} environment is lesser than the \textit{stable} environment, because more number of resubmissions are needed in the \textit{normal} environment.
The resource waste is higher in \textit{ReplicateAll(3)} than CRCH by 70\%, 47\% and 29\% for the \textit{stable}, \textit{normal} and \textit{unstable} environments respectively.
Overall CRCH performs better than its counterparts.

\begin{figure*}[!ht]
    \centering
\includegraphics[width=0.75\textwidth]{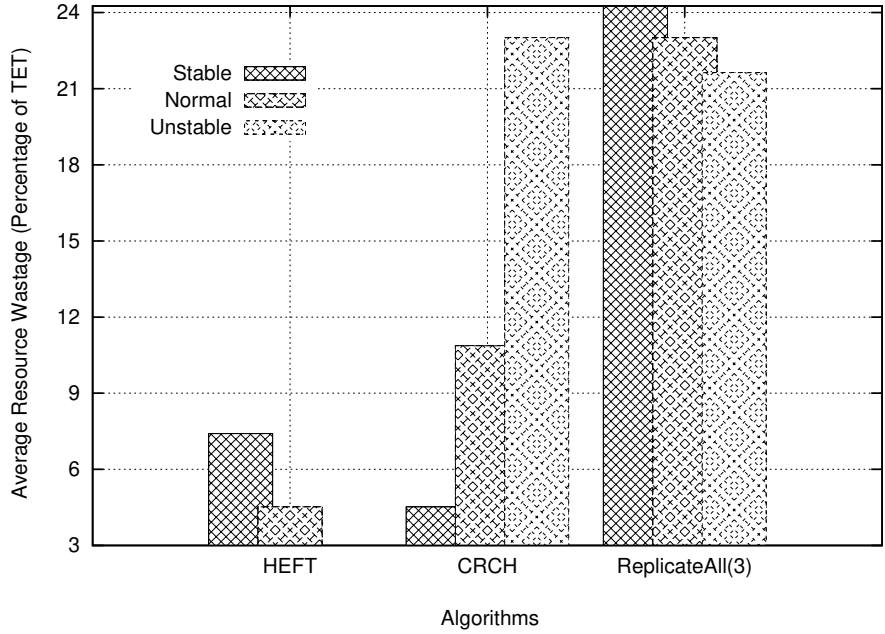}
    \caption{Average Resource Wastage}
    \label{Average Resource Wastage}
\end{figure*}

\begin{figure*}[!ht]
\centering
\includegraphics[width=0.75\textwidth]{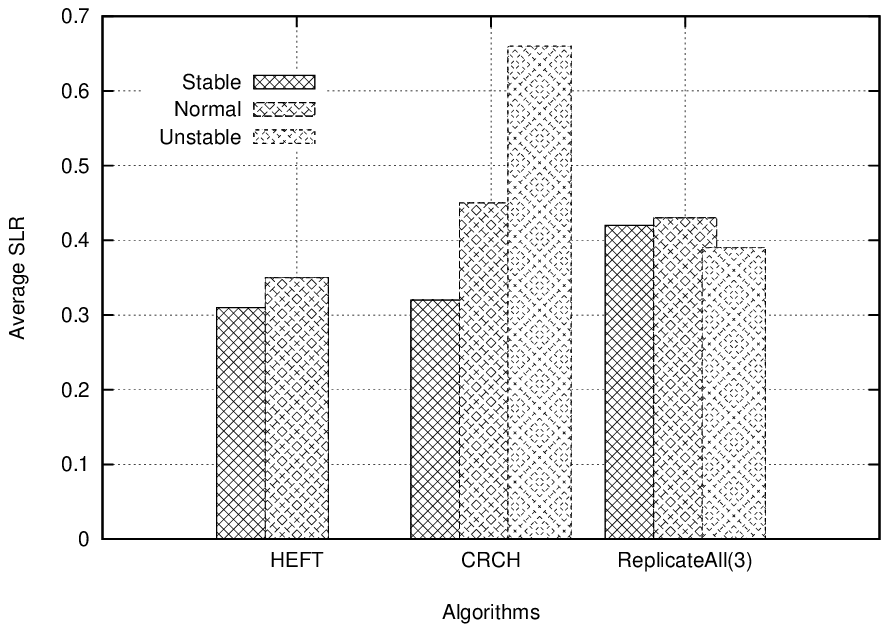}
\caption {Average SLR}
\label{SLR}
\end{figure*}

\par SLR is defined as the ratio between the execution time and the B-level of the first entry task of the workflow \cite{zhang2009combined}. In the case of CRCH and \textit{ReplicateAll(3)}, SLR is calculated as the ratio between the execution time and the B-level of the first task on the critical path after introducing the replicas. 
The B-level of a node $n_{i}$  is the length of the longest path from node $n_{i}$ to an exit node bound by the length of the critical path \cite{yukwong1999}. An optimal algorithm tries to choose the best scheduling plan and thus lowers the SLR value. HEFT has no consideration for task or resource failures and hence has low SLR value. CRCH assumes resource failures, and therefore the optimal resource for a task may not be available at the ready time of the task. The task might need to be scheduled on a second-best resource or wait for the failed resource to come online, and thus the SLR values are higher than HEFT. In the \textit{stable} environment, as shown in Figure \ref{SLR}, the SLR values of CRCH are comparable with that of HEFT and is only slightly higher by 5\%. The slight improvement in SLR for CRCH can be intuitively attributed to the reason that SLR is dependent on the critical path identified. Since every task has some number of replicas, a shorter critical path may be found from a replica of the task instead of the original one. In the case of failure of an optimal resource in CRCH, resubmissions on the same resource improve the SLR value. In a \textit{normal} environment more resources are expected to fail, hence the increase in SLR value for CRCH. Both the execution time (numerator of SLR) and the critical path found (denominator of SLR) increase and hence the increase of 10\% is observed. In \textit{ReplicateAll(3)}, there is a slight increase in SLR, but not as substantial as both the numerator and denominator of the SLR ratio increase. SLR is also affected by resubmissions and there is no resubmission in \textit{ReplicateAll(3)}.

\par
\begin{figure*}[t]
    \centering
    \includegraphics[width=1.1\textwidth]{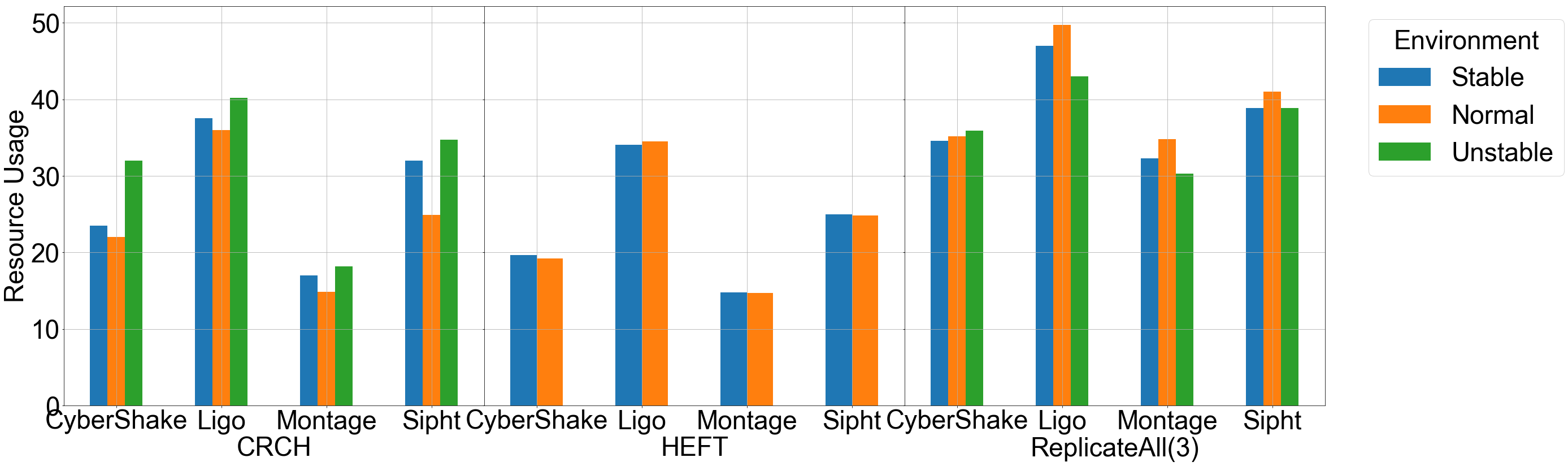}
    \caption{Resource Usage Across Workflow Types}
    \label{ResourceUsageAcrossTypes}
\end{figure*}

\begin{figure*}[!ht]
     \centering
    \includegraphics[width=1.1\textwidth]{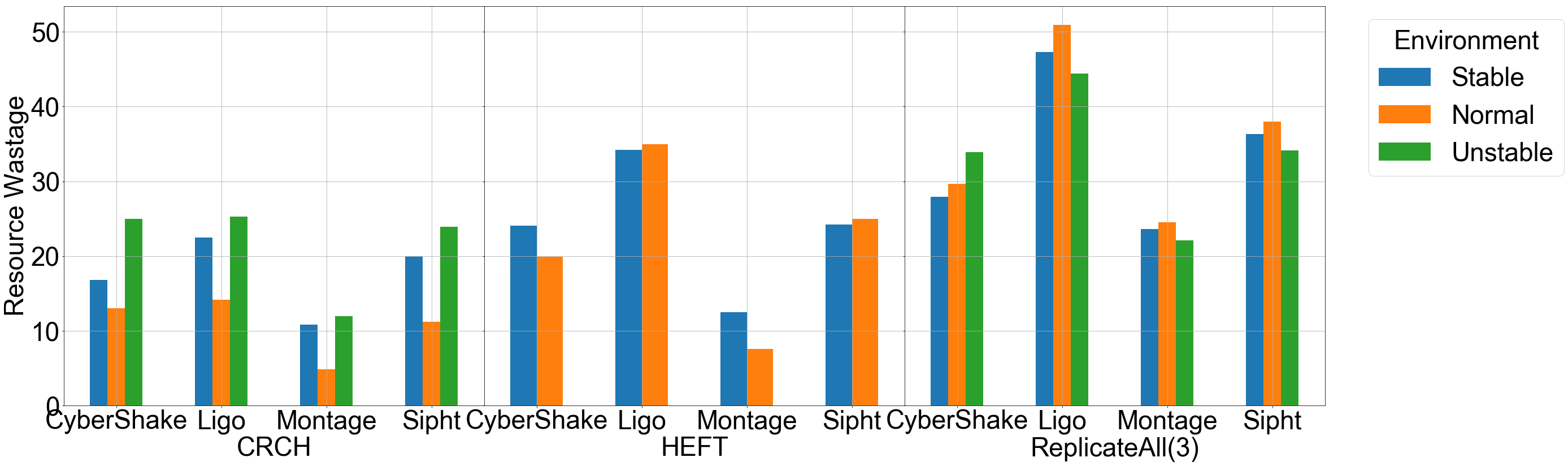}
    \caption{Resource Wastage Across Workflow Types}
    \label{ResourceWastageAcrossTypes}
\end{figure*}

 Metrics like Average Resource Usage (Figure \ref{ResourceUsageAcrossTypes}) and Average Resource Wastage (Figure \ref{ResourceWastageAcrossTypes}) show specific trends across workflow types which resonate with the structural differences, different data and computational characteristics among them. 
As CyberShake is much more CPU intensive than Montage
\cite{juve2012resource}, a 35\% increase in Resource Usage is seen under the normal environment with the CRCH Algorithm. This increase is as high as 129\% for LIGO (used in the physics field with the aim of detecting gravitational waves) which is characterized by having CPU intensive tasks that consume large memory \cite{juve2012resource}. Due to the fixed number of replicas, the Resource Usage is higher for \textit{ReplicateAll(3)}, as compared to CRCH under all environments and workflow types. It is interesting to note that while comparing each  workflow's Resource Usage to that of Montage with the \textit{ReplicateAll(3)} algorithm, the percentage increase is lower compared to CRCH. This is because the Resource Usage attributed to futile replication overshadows the Resource Usage attributed to inherent workflow characteristics. For example, there is a mere 47\% increase when LIGO is compared to Montage. An increase of 111\%, 129\%, and 144\% respectively is observed while comparing LIGO to Montage in the three different environments under CRCH. This rise in percentage is once again caused by the reduction in the failure rates of replicas as the environment stability improves. Identical patterns are observed while comparing Montage with CyberShake, and CyberShake with SIPHT.
\parskip 0pt
\par Corollary to the reasons mentioned for Resource Usage, similar trends are observed in the Average Resource Wastage. A 109\% increase is seen while comparing Montage with LIGO under the normal environment using the CRCH Algorithm. This figure is close to the 106\% increase seen in case of \textit{ReplicateAll(3)} as the number of unused replicas is comparable in most of the random draws. Montage has a high I/O and low CPU utilization \cite{juve2012resource} and hence has the least Resource Wastage irrespective of the algorithm or the environment. The reverse is true for SIPHT \cite{juve2012resource} and the increase in wastage is nominal. For a heavily CPU intensive workflow like LIGO \cite{juve2012resource} the percentage increase is as high as 211\% while comparing Montage with LIGO under the stable environment using the CRCH algorithm. This percentage increase drops to 110\% in the unstable environment owing to the drop in MTBF (or increase in failure rate). In other words, the unstable environment justifies the creation of more replicas and checkpoints.

\section{Conclusion and Future Work}
\label{conclusion}

It has been shown that modeling replication heuristics using unsupervised statistical techniques on readily available unlabelled data, improves efficiency in faulty environments. In other words, CRCH can capture the correlations between similar tasks and their corresponding replication counts. On the other hand, it fails to incorporate the probability distributions over resource failure parameters like MTTR and MTBF, in deciding the magnitude of task replication. This would mean that corresponding tasks in identical workflows would end up having a similar number of replications, irrespective of the environment of execution.

The CRCH analysis goes to show that using checkpoints to store light-weight pointers to saved states not only improves memory access time but also makes the system robust. Distributed non-volatile storage complements the use of checkpoints. The analysis reflects an improvement in resource wastage. Resources on an average spend less time recomputing intermediate results. Since the results are also stored at a non-volatile memory location, improving checkpoint retrieval through increased network capacity is trivial.


It has been proven that varying checkpoint parameters with failure parameters, can lead to better Average Total Execution Time. Albeit, the resource failure distributions for different resources are assumed to be mutually independent over time. This may not be the case when resource failures are modeled at a resource group level. In such situations, the probability of failure of a resource is dependent on the current state of its neighborhood resources and the resource failure probability distributions are merely conditionally independent, given the state of a few other resources \cite{poola2014robust}.

 The improvements in Total Execution Time, Resource Usage and Resource Wastage, over other HEFT based algorithms across normal, stable and unstable environments has been shown empirically. Further work can be done by establishing theoretical bounds on the extent of improvements in the aforementioned metrics.
 
   The failure models considered assume the availability of a few reliable resources, which are used for task resubmissions, ensuring that a task is resubmitted at most once. Thus helping in estimating the expected resubmission overhead. Although in practical circumstances, reliable resources may not always be available.

%
\par This paper can be extended to deal with real-world workflow management systems like Pegasus \cite{deelman2015pegasus}. An elaborate set of training samples for replication counts can further improve the machine learning aspect. The DAX files for tasks with multiple direct and derived features can also improve the heuristic accuracy. 

In estimating the increase in Total Execution Time, the model takes into account only the failures of tasks that lie on the critical path as they have a direct impact on the increase in TET. The DAX inputs with multiple/dynamic critical paths or cloud environments with highly varying MTBF/MTTR distributions for individual VMs requires further analysis.




\newpage
\section*{References}
\Urlmuskip=0mu plus 1mu\relax
\bibliographystyle{elsarticle-num}
\bibliography{mybibfile1}   



\end{document}